\documentclass[aps,prl,twocolumn,superscriptaddress,citeautoscript]{revtex4}

\usepackage{graphicx}
\usepackage{amsmath}

\graphicspath{{.}{./EPS/}}

\begin{document}

\title{Strongly Correlated Fractional Quantum Hall Line Junctions}

\author{U. Z\"ulicke}
\affiliation{Institut f\"ur Theoretische Festk\"orperphysik,
Universit\"at Karlsruhe, D-76128 Karlsruhe, Germany}

\author{E. Shimshoni}
\affiliation{Department of Mathematics--Physics, 
University of Haifa at Oranim, Tivon 36006, Israel}
\affiliation{Department of Physics and Astronomy, Rutgers University,
Piscataway, NJ 08854--8019}

\date{\today}

\begin{abstract}
We have studied a clean finite--length line junction between
interacting counterpropagating single--branch
fractional--quantum--Hall edge channels. Exact solutions for
low--lying excitations and transport properties are obtained when the
two edges belong to quantum Hall systems with {\em different\/}
filling factors and interact via the long--range Coulomb interaction.
Charging effects due to the coupling to external edge--channel leads
are fully taken into account. Conductances and power laws in the
current--voltage characteristics of tunneling are strongly affected by
inter--edge correlations.
\end{abstract}

\pacs{73.43.Cd, 73.43.Jn}

\maketitle

Two--dimensional (2D) electron systems exhibit incompressibilities
when their sheet density $n_0$ is commensurate with the value $B$ of
perpendicular magnetic field such that the filling factor $\nu=2\pi
\hbar n_0/|e B|$ is integer or equal to certain
fractions~\cite{qhe-sg,qhepersp}. Such an incompressible phase is
characterized by a quantized value of the Hall resistance and
low--lying excitations that are localized at the boundary of the 2D
system~\cite{bih:prb:82,pav:prb:84,ahm:prl:90,wen:prb:90,
froh:nuclB:91}. When $\nu$ is the inverse of an odd integer, these
quantum--Hall (QH) edge excitations have been shown~\cite{wen:int:92}
to be isomorphous to those of a single--branch chiral one--dimensional
(1D) electron system~\cite{voit:reprog:94}. Similar to their nonchiral
counterparts that are realized, e.g., in semiconductor quantum
wires~\cite{yac:prl:96} or carbon nanotubes~\cite{nanotube}, QH edges
are expected~\cite{wen:int:92} to exhibit power laws in electronic
correlation functions that are the hallmark of {\em Luttinger--liquid\/}
behavior~\cite{fdmh:jpc:81}. Virtually no other quasi--1D electron
system, however, matches the versatility of QH edges in tailoring
their electronic properties which can be achieved, e.g., simply by
adjusting the magnetic field and/or appropriate nanostructuring
techniques. Recent applications of the {\em cleaved--edge overgrowth\/}
method~\cite{cleaved1,cleaved2} have succeeded in creating extended
uniform tunnel junctions between two integer QH
edges~\cite{kang:nat:99} or a 2D electron system and a QH
edge~\cite{michi:physe:01}. Besides opening up new possibilities for
electron tunneling spectroscopy~\cite{uz:prb-rc:02b}, these new sample
geometries enable the controlled experimental study of the interplay
between tunneling and interaction effects in low
dimensions~\cite{fink:prb:93,fab:prb:93,pwa:prl:94,hjs:prb-rc:96} that
cannot be performed in conventional systems.

A diverse set of electron--correlation effects has been proposed for
QH line junctions where both edge channels belong to QH systems having
{\em the same\/} filling factor\cite{qhlinej1,qhlinej2,aditi:prb-rc:01,
kollar:prb-rc:02,fradline}. In our work presented here, we consider
the case of a {\em clean\/} junction where the filling factors
$\nu_{\text{R}}$ and $\nu_{\text{L}}$, characterizing the respective
right--moving and left--moving edges, are {\em different}. Such a
situation could be realized, e.g., in samples with two mutually
perpendicular 2D electron systems~\cite{michi:physe:01} and a magnetic
field properly adjusted in magnitude and direction. In the
experimentally realistic limit where electrons from the two edges
interact strongly via Coulomb interactions, low--lying excitations are
represented by two decoupled bosonic modes that have opposite
chirality. One of them, the charged mode, corresponds to fluctuations
in the total electron density in the two edges and is free. The
dynamics of its orthogonal complement, the neutral mode, turns out to
be governed by realizations of chiral sine--Gordon
models~\cite{uz:prl:99,naud:nucl:00} that can be solved {\em exactly\/}
for certain values of the parameter $\tilde\nu=\nu_{\text{R}}
\nu_{\text{L}}/|\nu_{\text{R}}-\nu_{\text{L}}|$. We use these exact
solutions to calculate transport through the line junction and find
that the strong inter--edge correlations significantly affect the
conductance and tunneling resonances arising from translational
invariance along the barrier.

\begin{figure}[b]
\includegraphics[width=2in]{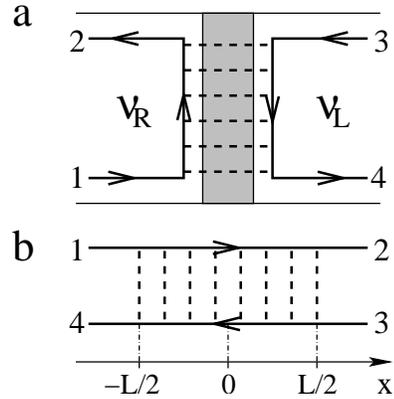}
\caption{Schematic layout of a quantum--Hall line junction. An
extended uniform tunnel barrier couples counterpropagating
single--branch edge channels from two 2D electron systems that have
different fractional filling factors $\nu_{\text{R}}$ and
$\nu_{\text{L}}$. (See panel~a.) We model this situation by two
parallel infinite chiral fractional edge channels that are coupled via
uniform tunneling and Coulomb interactions along the finite junction
region where $|x|\le L/2$. Outside the junction, edge branches
correspond to leads connecting to external reservoirs.
\label{sketch}}
\end{figure}

Starting point of our calculations is the Hamiltonian for our model of
the line junction. (See Fig.~\ref{sketch}.) It reads $H_{\text{J}}=
\int_{-L/2}^{L/2} dx\,\left({\mathcal H}_{\text{R}}+{\mathcal
H}_{\text{L}}+{\mathcal H}_{\text{int}}+{\mathcal H}_{\text{tun}}
\right)$ where ${\mathcal H}_{\text{R/L}}$ describe isolated edge
channels, ${\mathcal H}_{\text{int}}$ the inter--edge interaction, and
${\mathcal H}_{\text{tun}}$ uniform tunneling across the barrier:
\begin{subequations}
\begin{eqnarray}
{\mathcal H}_{\text{R/L}}&=&\frac{1}{4\pi}\left\{\hbar
v_{\text{R/L}}^{(0)}+\frac{\nu_{\text{R/L}}}{2\pi}\, U\right\}\,\,
\left[\partial_x\phi_{\text{R/L}}\right]^2\,\, , \\
{\mathcal H}_{\text{int}} &=& \frac{\sqrt{\nu_{\text{R}}\nu_{\text{L}}
}}{4\pi^2}\,\lambda\, U\,\, \partial_x\phi_{\text{R}}\,\, \partial_x
\phi_{\text{L}} \quad ,\\
{\mathcal H}_{\text{tun}} &=& t \left\{\psi_{\text{R}}^\dagger(x)
\psi_{\text{L}}(x)+ \text{H.c.}\right\}\quad . 
\end{eqnarray}
\end{subequations}
Here we denote the second--quantized annihilation operators for
electrons from the two edges as $\psi_{\text{R/L}}$. The chiral
bosonic phase fields $\phi_{\text{R/L}}(x)$ are related to the
respective edge densities via $\varrho_{\text{R/L}}=
\psi_{\text{R/L}}^\dagger\psi_{\text{R/L}}=\sqrt{\nu_{\text{R/L}}}\,
\partial_x\phi_{\text{R/L}}/(2\pi)$ and obey the commutation relations
$\big[\phi_{\text{R/L}}(x),\phi_{\text{R/L}}(x^\prime)\big]=\pm i\pi\,
\text{sgn}(x-x^\prime)$. The bare (confinement--induced) electron
velocities in the two edge channels are given by $v_{\text{R/L}}^{(0)}
$, $U$ is the matrix element of screened Coulomb interactions, and we
have taken into account the generally different interaction strengths
within and between the edges by a factor $0<\lambda\le 1$. It is
straightforward to diagonalize the part $H_{\text{J}}-
\int_{-L/2}^{L/2} dx\,\, {\mathcal H}_{\text{tun}}$, and using the
bosonization identity~\cite{vondelft}
\begin{equation}\label{bosid}
\psi_{\text{R/L}}(x)=\sqrt{z_{\text{R/L}}}\,\,{\mathcal
F}_{\text{R/L}}\,\,e^{i x \frac{Y_{\text{R/L}}}{\ell^2} \pm i
\frac{\phi_{\text{R/L}}(x)}{(\nu_{\text{R/L}})^{1/2}}},
\end{equation}
the tunneling term ${\mathcal H}_{\text{tun}}$ can be expressed solely
in terms of bosonic phase fields as well. [In Eq.~(\ref{bosid}),
$z_{\text{R/L}}$ are normalization constants, ${\mathcal
F}_{\text{R/L}}$ Klein factors, $Y_{\text{R/L}}$ the guiding--center
coordinates of edge electrons in the direction perpendicular to the
junction, and $\ell=\sqrt{\hbar c/|e B|}$ the magnetic length.]
However, explicit expressions for normal modes in the most general
case are unilluminating. We therefore proceed immediately to the
special case of interest to us where strong Coulomb interactions
dominate the bare electron velocities in each branch, i.e.,
$\hbar v_{\text{R/L}}^{(0)}\ll U$, and the distance between the two
edges is of the order of or smaller than $\ell$ such that $\lambda\to 
1$. It is then possible to rewrite the QH--junction Hamiltonian as
$H_{\text{J}}=\int_{-L/2}^{L/2} dx\,\left({\mathcal H}_{\text{c}}+
{\mathcal H}_{\text{n}}\right)$ with the contributions
\begin{subequations}\label{cnham}
\begin{eqnarray}
{\mathcal H}_{\text{c}}&=&\frac{\hbar v_{\text{c}}}{4\pi}\,\left[
\partial_x\phi_{\text{c}}\right]^2\quad , \\ \label{neutham}
{\mathcal H}_{\text{n}}&=&\frac{\hbar v_{\text{n}}}{4\pi}\,\left[
\partial_x\phi_{\text{n}}\right]^2 +2 t\sqrt{z_{\text{R}}z_{\text{L}}}
\,\cos\left(\frac{\phi_{\text{n}}}{\sqrt{\tilde\nu}} + \frac{x \Delta}
{\ell^2}\right) ,
\end{eqnarray}
\end{subequations}
for independent c(harged) and n(eutral) modes~\footnote{This actually
requires also the condition $U\gg |t|\ell$ to hold which is
typically satisfied. No Klein factors appear in Eqs.~(\ref{cnham}) as
they give rise only to an overall factor $\pm 1$ that can be absorbed
into the tunneling matrix element~\cite{naud:nucl:00}.}.
Their corresponding phase fields obey commutation relations $\big[
\phi_{\text{c/n}}(x),\phi_{\text{c/n}}(x^\prime)\big]=\pm\,\text{sgn}
(\nu_{\text{R}}-\nu_{\text{L}})\, i\pi\,\text{sgn}(x-x^\prime)$ and
are related to the densities of the original right--moving and
left--moving edge electrons via
\begin{equation}
\varrho_{\text{R/L}}=\frac{\pm\text{sgn}(\nu_{\text{R}}-\nu_{\text{L}}
)}{2\pi\sqrt{|\nu_{\text{R}}-\nu_{\text{L}}|}}\left(\nu_{\text{R/L}}\,
\partial_x\phi_{\text{c}} - \sqrt{\nu_{\text{R}}\nu_{\text{L}}}\,
\partial_x\phi_{\text{n}}\right)\, .
\end{equation}
The velocities are $v_{\text{c}}=|\nu_{\text{R}}-\nu_{\text{L}}|\, U/(
2\pi\hbar)$ and $v_{\text{n}}=\big(\nu_{\text{R}}v_{\text{L}}^{(0)}+
\nu_{\text{L}}v_{\text{R}}^{(0)}\big)/|\nu_{\text{R}}-\nu_{\text{L}}|
$, and we used the abbreviation $\Delta :=Y_{\text{R}}-Y_{\text{L}}$.
We see from Eqs.~(\ref{cnham}) that the charged mode is free while the
neutral--mode dynamics is that of a chiral sine--Gordon
model~\cite{uz:prl:99,naud:nucl:00}. Before obtaining its solution, we
interject an elaboration on observables related to electron transport.

To discuss transport properties, it is necessary to consider
continuity equations and chemical potentials for the various edge
densities. Denoting time by $\tau$, the operator for particle current
(per unit length) across the barrier is
\begin{subequations}
\begin{eqnarray}
{\mathcal I}_{\text{J}}(x) &=& -\frac{d}{d\tau}\,\varrho_{\text{R}}(x)
= \frac{d}{d\tau}\,\varrho_{\text{L}}(x) \quad , \\
&=& i\, \frac{t}{\hbar}\,\left\{\psi_{\text{R}}^\dagger(x)
\psi_{\text{L}}(x) - \text{H.c.}\right\}\quad ,
\end{eqnarray}
\end{subequations}
In our limit of interest, its bosonized expression depends on the
neutral mode only,
\begin{equation}\label{neutcurr}
{\mathcal I}_{\text{J}}(x) = 2\, \frac{t}{\hbar}\,\sqrt{z_{\text{R}}
z_{\text{L}}}\,\sin\left(\frac{\phi_{\text{n}}}{\sqrt{\tilde\nu}} +
\frac{x \Delta}{\ell^2}\right)\,\, .
\end{equation}
It is straightforward to find a continuity equation for density
$\varrho_{\text{c}}:=\partial_x\phi_{\text{c}}/(2\pi)$ associated with
the charged mode,
\begin{subequations}
\begin{eqnarray}
0 &=& \frac{d}{d\tau}\,\frac{\varrho_{\text{R}}+\varrho_{\text{L}}}
{\sqrt{|\nu_{\text{R}}-\nu_{\text{L}}|}} \equiv  \frac{d}{d\tau}\,
\varrho_{\text{c}} \quad , \\
  &=& \left\{\partial_\tau+\text{sgn}(\nu_{\text{R}}-\nu_{\text{L}})
\, v_{\text{c}}\,\partial_x\right\}\varrho_{\text{c}} \quad ,
\end{eqnarray}
\end{subequations}
and the corresponding one for $\varrho_{\text{n}}:=\partial_x
\phi_{\text{n}}/(2\pi)$,
\begin{subequations}
\begin{eqnarray}
0 &=& \frac{d}{d\tau}\,\varrho_{\text{n}} - \frac{\text{sgn}(
\nu_{\text{R}}-\nu_{\text{L}})}{\sqrt{\tilde\nu}}\,{\mathcal
I}_{\text{J}}\quad , \\
  &=& \left\{\partial_\tau-\text{sgn}(\nu_{\text{R}}-\nu_{\text{L}})
\, v_{\text{n}}\,\partial_x\right\}\varrho_{\text{n}} \quad .
\end{eqnarray}
\end{subequations}
In the stationary regime, densities have no explicit time dependence.
The above continuity equations then imply that they assume constant
values $\bar\varrho_{\text{c/n}}$ along the junction. We can also
derive expressions for local chemical potentials $\mu_{\text{c/n}}:=
\delta H_{\text{J}}/\delta\varrho_{\text{c/n}}\equiv \mp i\,\text{sgn}
(\nu_{\text{R}}-\nu_{\text{L}})\,\big[ H_{\text{J}}, \phi_{\text{c/n}}
\big]$ associated with the charged and neutral modes:
\begin{subequations}
\begin{eqnarray}
\mu_{\text{c}}(x) &=& 2\pi\hbar\, v_{\text{c}}\, \varrho_{\text{c}}(x)
\quad , \\ \label{neutchem}
\mu_{\text{n}}(x) &=& 2\pi\hbar\, v_{\text{n}}\, \varrho_{\text{n}}(x)
\nonumber \\ && + \frac{\pi\hbar}{\sqrt{\tilde\nu}}
\int_{-\frac{L}{2}}^{\frac{L}{2}} dx^\prime\,\,{\mathcal I}_{\text{J}}
(x^\prime)\,\,\text{sgn}(x - x^\prime)\, .
\end{eqnarray}
\end{subequations}
Specializing these to the stationary limit, we find
\begin{subequations}
\begin{eqnarray}
\mu_{\text{c}}(x) &=& 2\pi\hbar\, v_{\text{c}}\,\bar\varrho_{\text{c}}
= \text{const.}\quad , \\
\mu_{\text{n}}(\pm L/2) &=& 2\pi\hbar\, v_{\text{n}}\, \bar
\varrho_{\text{n}}\pm\frac{\pi\hbar}{\sqrt{\tilde\nu}}\, I_{\text{J}}
\quad ,
\end{eqnarray}
\end{subequations}
where $I_{\text{J}}:=\int_{-L/2}^{L/2}dx\,\,{\mathcal I}_{\text{J}}(x)
$ is the total current flowing through the junction. The values of
$\bar\varrho_{\text{c/n}}$ and $I_{\text{J}}$ can be related to
chemical potentials in the external leads that connect the QH junction
to reservoirs (see Fig.~\ref{sketch}), which we proceed to show now.

In a real setup as sketched in Fig.~\ref{sketch}a, edge channels
{\em away} from the junction are not coupled via tunneling or
interactions anymore. However, for each of these chiral 1D leads,
Coulomb interactions between electrons {\em at the same edge\/} still
give the dominant contribution to the edge--magnetoplasmon
velocity~\cite{uz:prb:96}. We therefore model the dynamics of
electrons in regions $|x|>L/2$ by the lead Hamiltonian $H_{\text{E}}=
\Big\{\int_{-\infty}^{-L/2}+\int_{L/2}^{\infty}\Big\}\, dx\,\left(
{\mathcal H}_{\text{R}}+{\mathcal H}_{\text{L}}\right)$. While
$H_{\text{E}}$ is not diagonal in the charged and neutral modes, their
corresponding chemical potentials are still well--defined in the lead
regions. As the densities in each of the four chiral edge--channel
leads cannot change, chemical potentials remain constant as
well~\cite{frohlich}. Assuming again $U\gg\hbar v_{\text{R/L}}^{(0)}$,
we find for these (in a compact notation where $\mu_{i/j}$ means
$\mu_i$ or $\mu_j$, respectively)
\begin{subequations}
\begin{eqnarray}
\left.\mu_{\text{c}}\right|_{x\stackrel{<}{>}\mp\frac{L}{2}}
&=& \frac{\nu_{\text{R}}\,\mu_{1/2} - \nu_{\text{L}}\,\mu_{4/3}}
{\sqrt{|\nu_{\text{R}}-\nu_{\text{L}}|}}\,\text{sgn}(\nu_{\text{R}}-
\nu_{\text{L}})\, , \\
\left.\mu_{\text{n}}\right|_{x\stackrel{<}{>}\mp\frac{L}{2}}
&=& -\sqrt{\tilde\nu}\left(\mu_{1/2}-\mu_{4/3}\right)\,\text{sgn}(
\nu_{\text{R}}-\nu_{\text{L}}) .
\end{eqnarray}
\end{subequations}
Demanding continuity of $\mu_{\text{c/n}}$ at $x=\pm L/2$ yields four
linear relations involving the chemical potentials in the four leads,
the constants $\bar\varrho_{\text{c/n}}$, and $I_{\text{J}}$. After
some straightforward algebra, we obtain the expressions
\begin{subequations}
\begin{eqnarray}
\mu_{2/4} &=& \mu_{1/3} \pm \frac{2\pi\hbar}{\nu_{\text{R/L}}}
\, I_{\text{J}}\quad , \\ \label{selfc2}
v_{\text{c}}\,\bar\varrho_{\text{c}}&=&\frac{\nu_{\text{R}}\mu_1 -
\nu_{\text{L}}\mu_3 + 2\pi\hbar\, I_{\text{J}}}{2\pi\hbar
\sqrt{|\nu_{\text{R}}-\nu_{\text{L}}|}}\text{sgn}(\nu_{\text{R}}-
\nu_{\text{L}})\, , \\ \label{selfconist}
v_{\text{n}}\,\bar\varrho_{\text{n}}&=&\frac{\mu_1 - \mu_3 + \left(
\frac{\pi\hbar}{\nu_{\text{R}}}+\frac{\pi\hbar}{\nu_{\text{L}}}\right)
\,I_{\text{J}}}{-2\pi\hbar/\sqrt{\tilde\nu}}\text{sgn}(\nu_{\text{R}}
-\nu_{\text{L}}) ,
\end{eqnarray}
\end{subequations}
relating the `output' chemical potentials $\mu_{2/4}$ as well as the
constants $\bar\varrho_{\text{c/n}}$ to the experimentally adjustable
`input' chemical potentials $\mu_{1/3}$ and the {\it a priori\/}
unknown current $I_{\text{J}}$~\footnote{The result that external
voltages enter as {\em uniform\/} constraints on chiral charge densities
{\em in the tunneling region\/} is nontrivial. Similar relations arise for a
homogeneous interacting quantum wire {\em in the absence of backscattering}.
See, e.g., R. Egger and H. Grabert, Phys. Rev. B {\bf 58}, 10761
(1998); I. Safi, Eur. Phys. J. B {\bf 12}, 451 (1999).}. To obtain the
full solution for the transport problem of our finite QH junction that
is attached to edge--channel leads, we are now left with the task to
determine $\bar\varrho_{\text{n}}$ and $I_{\text{J}}$
self--consistently using the dynamics of the neutral mode expressed in
Eqs.~(\ref{neutham}), (\ref{neutcurr}), (\ref{neutchem}), and
(\ref{selfconist}). We now show how this can be achieved.

The Hamiltonian of the neutral mode [given by Eq.~(\ref{neutham})] is
a realization of recently discussed~\cite{uz:prl:99,naud:nucl:00}
{\em chiral\/} sine--Gordon models. In contrast to the well--known
nonchiral sine--Gordon model~\cite{tsvelik}, the quantum character of
its dynamics arises because the bosonic field entering the cosine term
does not commute with itself. Possible values of the parameter $\tilde
\nu$ for the QH line junction discussed here can be parameterized by
two non--negative integers $m, m^\prime$ such that $1/\tilde\nu=2|m-
m^\prime|$. These values correspond to QH junctions between systems
having filling factors $1/(2m+1)$ and $1/(2m^\prime+1)$, respectively.
It has been shown~\cite{naud:nucl:00} that the cosine term constitutes
only an irrelevant perturbation to the dynamics of the chiral bosonic
field when $|m-m^\prime|>2$. More interesting, and easier to realize
experimentally, are the two cases $1/\tilde\nu\in\{2, 4\}$ where the
cosine term turns out to be relevant or marginal, respectively, and
exact solutions for its dynamics can be
found~\cite{uz:prl:99,naud:nucl:00}.

Lets first consider the case $\tilde\nu=1/2$ realized, e.g., at a line
junction between QH systems having filling factors $1$ and $1/3$,
respectively. Introducing an {\em auxiliary\/} chiral boson field
$\eta(x)$ that has the same chirality and velocity as $\phi_{\text{n}}
$, we can define fictitious chiral fermion operators
$\Psi_{\uparrow/\downarrow}$ via an inverse bosonization identity
$\Psi_\sigma\propto\exp\left\{-i\,\text{sgn}(\nu_{\text{R}}-
\nu_{\text{L}})\left[\eta+\sigma\phi_{\text{n}}\right]/\sqrt{2}\right
\}$. Note that the pseudospin degree of freedom indexed by $\sigma$ is
not related to the spin of the original electrons in the sample. In
the new fictitious--fermion degrees of freedom, Eqs.~(\ref{neutham})
and (\ref{neutcurr}) are quadratic:
\begin{subequations}\label{efftunn}
\begin{eqnarray}
{\mathcal H}_{\text{n}}^\prime&=&\sum_\sigma\Psi_\sigma^\dagger\left[
i\,\text{sgn}(\nu_{\text{R}}-\nu_{\text{L}})\, \hbar v_{\text{n}}
\partial_x\right]\Psi_\sigma \nonumber\\
&&\hspace{0.5cm} +\, t\,\left(\Psi_\uparrow^\dagger\Psi_\downarrow
e^{i\frac{x\Delta}{\ell^2}\,\text{sgn}(\nu_{\text{R}}-\nu_{\text{L}})}
+ \text{H.c.}\right)\quad , \\
{\mathcal I}_{\text{J}}^\prime&=&-i\, \frac{t}{\hbar}\, \left(
\Psi_\uparrow^\dagger \Psi_\downarrow e^{i\frac{x\Delta}{\ell^2}\,
\text{sgn}(\nu_{\text{R}}-\nu_{\text{L}})} - \text{H.c.}\right)\quad .
\end{eqnarray}
\end{subequations}
In addition, density and chemical potential of the neutral mode are
related to corresponding fictitious--fermion quantities via
$\Psi_\uparrow^\dagger\Psi_\uparrow - \Psi_\downarrow^\dagger
\Psi_\downarrow\equiv\sqrt{2}\,\varrho_{\text{n}}$ and $\mu_\uparrow-
\mu_\downarrow\equiv\sqrt{2}\,\mu_{\text{n}}$. Note that only the
pseudospin sector of the new fermionic theory has any bearing for real
observations, while the pseudo{\em charge\/} sector described by the
auxiliary field $\eta$ is hidden. We see that the problem of the
original line junction between two strongly interacting edge branches
with opposite chirality has been mapped onto that of tunneling between
two branches of {\em noninteracting\/} and {\em chiral\/} fermions. The latter
is readily solved using standard methods. For example, when tunneling
is weak enough such that it is possible to neglect the contribution
$\propto I_{\text{J}}$ in Eq.~(\ref{selfconist}), we find $e^2
I_{\text{J}}=G_{\text{J}}\,(\mu_1-\mu_3)$ with the linear electric
conductance 
\begin{equation}
G_{\text{J}} = \frac{e^2}{2\pi\hbar}\,\frac{\sin^2\left[\frac{\pi L}
{L_t}\sqrt{1+\xi^2}\right]}{1 + \xi^2} \quad .
\end{equation}
Here $L_t=\pi\hbar v_{\text{n}}/|t|$ is a fundamental length scale set
by tunneling, and $\xi=\hbar v_{\text{n}}\Delta/(2|t|\ell^2)$ a
resonance parameter: $G_{\text{J}}$ is maximal for $\xi\to 0$ which
corresponds to the resonance condition where both energy and 1D
momentum are conserved in a tunneling event~\footnote{Note that the
condition of weak tunneling used here is valid for $G_{\text{J}}\ll
e^2/2\pi\hbar$, yet it does not restrict the tunneling strength to the
perturbative limit $L\ll L_t$.}. The obtained {\em linear\/} dependence
of $I_{\text{J}}$ on $\mu_1-\mu_3$ has to be contrasted with the power
law $I_{\text{J}}\propto(\mu_1-\mu_3)^{\nu_{\text{R}}^{-1}+
\nu_{\text{L}}^{-1}-1}$ that is to be expected for momentum--resolved
tunneling at a line junction between {\em noninteracting\/}
chiral--Luttinger--liquid~\cite{wen:int:92} edge channels.
Furthermore, chirality of the effective tunneling problem described by
Eqs.~(\ref{efftunn}) results in charge oscillations along the junction
similar to those predicted for QH bilayer systems~\cite{naud:nucl:00}
and parallel quantum wires~\cite{uz:prb:01}.

Before concluding, we discuss the other nontrivial case $\tilde\nu=1/
4$ which is realized, e.g., when $\nu_{\text{R}}=1$ and
$\nu_{\text{L}}=1/5$. Again it is possible to solve the problem by
refermionization~\cite{naud:nucl:00}. Defining $\psi_{\text{n}}(x)=
\sqrt{z_{\text{n}}}\,{\mathcal F}_{\text{n}}\exp\{-i\,\text{sgn}(
\nu_{\text{R}}-\nu_{\text{L}})\left[\phi_{\text{n}}(x) + x\Delta/2
\ell^2\right]\}$ and using the identity~\cite{naud:nucl:00}
$\psi_{\text{n}}(x)i\,\partial_x\psi_{\text{n}}(x)=2\pi\,\text{sgn}(
\nu_{\text{R}}-\nu_{\text{L}})z_{\text{n}}^2{\mathcal F}_{\text{n}}^2
\exp\{-[2 i \phi_{\text{n}}(x) + i x \Delta/\ell^2]\}$, we can rewrite
Eqs.~(\ref{neutham}) and (\ref{neutcurr}) as quadratic forms in the
fictitious chiral Dirac fermion $\psi_{\text{n}}$. It turns out to be
useful to utilize its decomposition in terms of Majorana fermions,
$\psi_{\text{n}}=(\chi_+ + i\,\chi_-)/\sqrt{2}$, which yields
\begin{subequations}
\begin{eqnarray}
{\mathcal H}_{\text{n}}^{\prime\prime} &=& \frac{1}{2}\sum_{r=\pm}
\chi_r(i\hbar v_r\partial_x)\chi_r - \frac{\hbar v_{\text{n}}\Delta}
{\ell^2}\, i\, \chi_+ \chi_- \quad , \\
{\mathcal I}_{\text{J}}^{\prime\prime} &=& \frac{-i\,\tilde t}{2\pi
\hbar}\,\text{sgn}(\nu_{\text{R}}-\nu_{\text{L}})\left(\chi_+
\partial_x \chi_- +\chi_-\partial_x \chi_+\right)  \quad ,
\end{eqnarray}
\end{subequations}
where $v_\pm=\text{sgn}(\nu_{\text{R}}-\nu_{\text{L}})\,\left[
v_{\text{n}}\pm\tilde t/\pi\right]$ with $\tilde t = t
\sqrt{z_{\text{R}}z_{\text{L}}}/z_{\text{n}}^2$. The Hamiltonian
${\mathcal H}_{\text{n}}^{\prime\prime}$ is easily diagonalized, and
the occupation--number distribution for Majorana fermions in
reciprocal space is fixed by the requirement $i\,\chi_+\chi_-=\bar
\rho_{\text{n}}$. The transport problem can then be solved exactly
again. A complete analysis is left to a later
publication~\cite{longv}; here we can only mention results for $\Delta
=0$ and in the limit of weak tunneling where it is possible to neglect
the contribution proportional to $I_{\text{J}}$ in
Eq.~(\ref{selfconist}). Quite different from the above considered case
of $\tilde\nu=1/2$, we find here that $I_{\text{J}}$ is {\em oscillating
in time} with period $2(\pi\hbar)^2 v_{\text{n}}/(\tilde t |\mu_1-
\mu_3|)$. A similar dephasing effect as observed~\cite{tinlunho} in QH
line junctions where $\nu_{\text{R}}=\nu_{\text{L}}=1$ leads to the
temporal decay of the oscillation amplitude.

In summary, we have obtained exact solutions for transport through
finite QH line junctions where the two edge channels belong
to systems having different filling factors $\nu_{\text{R}}$ and
$\nu_{\text{L}}$. Charging effects have been treated fully
self--consistently by imposing appropriate boundary conditions for
chemical potentials in the attached edge--channel leads. Strong
coupling of edge channels via Coulomb interactions in a junction with
$\tilde\nu\equiv\frac{\nu_{\text{R}}\nu_{\text{L}}}{|\nu_{\text{R}}-
\nu_{\text{L}}|}=1/2$ gives rise to a linear IV--characteristics for
tunneling, as opposed to the power law expected in the absence of
inter--edge correlations. At junctions with $\tilde\nu=1/4$, tunneling
currents oscillate in time.

We thank N. Andrei, C. Chamon and V. Tripathi for useful discussions. 
This work was supported in part by the German Science Foundation (DFG)
through Grant No.~ZU~116/1 and the German Federal Ministry of
Education and Research (BMBF) in the framework of the DIP program.


\begin{thebibliography}{33}
\expandafter\ifx\csname natexlab\endcsname\relax\def\natexlab#1{#1}\fi
\expandafter\ifx\csname bibnamefont\endcsname\relax
  \def\bibnamefont#1{#1}\fi
\expandafter\ifx\csname bibfnamefont\endcsname\relax
  \def\bibfnamefont#1{#1}\fi
\expandafter\ifx\csname citenamefont\endcsname\relax
  \def\citenamefont#1{#1}\fi
\expandafter\ifx\csname url\endcsname\relax
  \def\url#1{\texttt{#1}}\fi
\expandafter\ifx\csname urlprefix\endcsname\relax\def\urlprefix{URL }\fi
\providecommand{\bibinfo}[2]{#2}
\providecommand{\eprint}[2][]{\url{#2}}

\bibitem[{\citenamefont{Prange and Girvin}(1990)}]{qhe-sg}
\bibinfo{editor}{\bibfnamefont{R.~E.} \bibnamefont{Prange}} \bibnamefont{and}
  \bibinfo{editor}{\bibfnamefont{S.~M.} \bibnamefont{Girvin}}, eds.,
  \emph{\bibinfo{title}{The Quantum Hall Effect}}
  (\bibinfo{publisher}{Springer}, \bibinfo{address}{New York},
  \bibinfo{year}{1990}), \bibinfo{edition}{2nd} ed.

\bibitem[{\citenamefont{{Das Sarma} and Pinczuk}(1997)}]{qhepersp}
\bibinfo{editor}{\bibfnamefont{S.}~\bibnamefont{{Das Sarma}}} \bibnamefont{and}
  \bibinfo{editor}{\bibfnamefont{A.}~\bibnamefont{Pinczuk}}, eds.,
  \emph{\bibinfo{title}{Perspectives in the Quantum Hall Effects}}
  (\bibinfo{publisher}{Wiley}, \bibinfo{address}{New York},
  \bibinfo{year}{1997}).

\bibitem[{\citenamefont{Halperin}(1982)}]{bih:prb:82}
\bibinfo{author}{\bibfnamefont{B.~I.} \bibnamefont{Halperin}},
  \bibinfo{journal}{Phys. Rev. B} \textbf{\bibinfo{volume}{25}},
  \bibinfo{pages}{2185} (\bibinfo{year}{1982}).

\bibitem[{\citenamefont{MacDonald and St{\v r}eda}(1984)}]{pav:prb:84}
\bibinfo{author}{\bibfnamefont{A.~H.} \bibnamefont{MacDonald}}
  \bibnamefont{and} \bibinfo{author}{\bibfnamefont{P.}~\bibnamefont{St{\v
  r}eda}}, \bibinfo{journal}{Phys. Rev. B} \textbf{\bibinfo{volume}{29}},
  \bibinfo{pages}{1616} (\bibinfo{year}{1984}).

\bibitem[{\citenamefont{MacDonald}(1990)}]{ahm:prl:90}
\bibinfo{author}{\bibfnamefont{A.~H.} \bibnamefont{MacDonald}},
  \bibinfo{journal}{Phys. Rev. Lett.} \textbf{\bibinfo{volume}{64}},
  \bibinfo{pages}{220} (\bibinfo{year}{1990}).

\bibitem[{\citenamefont{Wen}(1990)}]{wen:prb:90}
\bibinfo{author}{\bibfnamefont{X.~G.} \bibnamefont{Wen}},
  \bibinfo{journal}{Phys. Rev. B} \textbf{\bibinfo{volume}{41}},
  \bibinfo{pages}{12838} (\bibinfo{year}{1990}).

\bibitem[{\citenamefont{Fr\"ohlich and Zee}(1991)}]{froh:nuclB:91}
\bibinfo{author}{\bibfnamefont{J.}~\bibnamefont{Fr\"ohlich}} \bibnamefont{and}
  \bibinfo{author}{\bibfnamefont{A.}~\bibnamefont{Zee}},
  \bibinfo{journal}{Nucl. Phys. B} \textbf{\bibinfo{volume}{B364}},
  \bibinfo{pages}{517} (\bibinfo{year}{1991}).

\bibitem[{\citenamefont{Wen}(1992)}]{wen:int:92}
\bibinfo{author}{\bibfnamefont{X.~G.} \bibnamefont{Wen}},
  \bibinfo{journal}{Int. J. Mod. Phys. B} \textbf{\bibinfo{volume}{6}},
  \bibinfo{pages}{1711} (\bibinfo{year}{1992}).

\bibitem[{\citenamefont{Voit}(1994)}]{voit:reprog:94}
\bibinfo{author}{\bibfnamefont{J.}~\bibnamefont{Voit}}, \bibinfo{journal}{Rep.
  Prog. Phys.} \textbf{\bibinfo{volume}{57}}, \bibinfo{pages}{977}
  (\bibinfo{year}{1994}).

\bibitem[{\citenamefont{Yacoby et~al.}(1996)\citenamefont{Yacoby, St\"ormer,
  Wingreen, Pfeiffer, Baldwin, and West}}]{yac:prl:96}
\bibinfo{author}{\bibfnamefont{A.}~\bibnamefont{Yacoby}}
  \bibnamefont{ \textit{et~al.}},
  \bibinfo{journal}{Phys. Rev. Lett.} \textbf{\bibinfo{volume}{77}},
  \bibinfo{pages}{4612} (\bibinfo{year}{1996}).

\bibitem[{\citenamefont{Wilder et~al.}(1998)\citenamefont{Wilder, Venema,
  Rinzler, Smalley, and Dekker}}]{nanotube}
\bibinfo{author}{\bibfnamefont{J.~W.~G.} \bibnamefont{Wilder}}
  \bibnamefont{ \textit{et~al.}},
  \bibinfo{journal}{Nature} \textbf{\bibinfo{volume}{391}}, \bibinfo{pages}{59
  } (\bibinfo{year}{1998}).

\bibitem[{\citenamefont{Haldane}(1981)}]{fdmh:jpc:81}
\bibinfo{author}{\bibfnamefont{F.~D.~M.} \bibnamefont{Haldane}},
  \bibinfo{journal}{J. Phys. C} \textbf{\bibinfo{volume}{14}},
  \bibinfo{pages}{2585} (\bibinfo{year}{1981}).

\bibitem[{\citenamefont{Pfeiffer et~al.}(1990)\citenamefont{Pfeiffer, West,
  St\"ormer, Eisenstein, Baldwin, Gershoni, and Spector}}]{cleaved1}
\bibinfo{author}{\bibfnamefont{L.}~\bibnamefont{Pfeiffer}}
  \bibnamefont{ \textit{et~al.}},
  \bibinfo{journal}{Appl. Phys. Lett.} \textbf{\bibinfo{volume}{56}},
  \bibinfo{pages}{1697} (\bibinfo{year}{1990}).

\bibitem[{\citenamefont{Grayson et~al.}(1996)\citenamefont{Grayson, Kurdak,
  Tsui, Parihar, Lyon, and Shayegan}}]{cleaved2}
\bibinfo{author}{\bibfnamefont{M.}~\bibnamefont{Grayson}}
  \bibnamefont{ \textit{et~al.}},
  \bibinfo{journal}{Solid-State Electron.} \textbf{\bibinfo{volume}{40}},
  \bibinfo{pages}{233} (\bibinfo{year}{1996}).

\bibitem[{\citenamefont{Kang et~al.}(1999)\citenamefont{Kang, Stormer,
  Pfeiffer, Baldwin, and West}}]{kang:nat:99}
\bibinfo{author}{\bibfnamefont{W.}~\bibnamefont{Kang}}
  \bibnamefont{ \textit{et~al.}},
  \bibinfo{journal}{Nature} \textbf{\bibinfo{volume}{403}}, \bibinfo{pages}{59}
  (\bibinfo{year}{1999}).

\bibitem[{\citenamefont{Huber et~al.}(2001)\citenamefont{Huber, Grayson,
  Rother, Deutschmann, Biberacher, Wegscheider, Bichler, and
  Abstreiter}}]{michi:physe:01}
\bibinfo{author}{\bibfnamefont{M.}~\bibnamefont{Huber}}
  \bibnamefont{ \textit{et~al.}},
  \bibinfo{journal}{Physica E} \textbf{\bibinfo{volume}{12}},
  \bibinfo{pages}{125} (\bibinfo{year}{2001}).

\bibitem[{\citenamefont{Z\"ulicke et~al.}(2002)\citenamefont{Z\"ulicke,
  Shimshoni, and Governale}}]{uz:prb-rc:02b}
\bibinfo{author}{\bibfnamefont{U.}~\bibnamefont{Z\"ulicke}},
  \bibinfo{author}{\bibfnamefont{E.}~\bibnamefont{Shimshoni}},
  \bibnamefont{and}
  \bibinfo{author}{\bibfnamefont{M.}~\bibnamefont{Governale}},
  \bibinfo{journal}{Phys. Rev. B} \textbf{\bibinfo{volume}{65}},
  \bibinfo{pages}{241315(R)} (\bibinfo{year}{2002}).

\bibitem[{\citenamefont{Finkel'stein and Larkin}(1993)}]{fink:prb:93}
\bibinfo{author}{\bibfnamefont{A.~M.} \bibnamefont{Finkel'stein}}
  \bibnamefont{and} \bibinfo{author}{\bibfnamefont{A.~I.}
  \bibnamefont{Larkin}}, \bibinfo{journal}{Phys. Rev. B}
  \textbf{\bibinfo{volume}{47}}, \bibinfo{pages}{10461} (\bibinfo{year}{1993}).

\bibitem[{\citenamefont{Fabrizio}(1993)}]{fab:prb:93}
\bibinfo{author}{\bibfnamefont{M.}~\bibnamefont{Fabrizio}},
  \bibinfo{journal}{Phys. Rev. B} \textbf{\bibinfo{volume}{48}},
  \bibinfo{pages}{15838} (\bibinfo{year}{1993}).

\bibitem[{\citenamefont{Clarke et~al.}(1994)\citenamefont{Clarke, Strong, and
  Anderson}}]{pwa:prl:94}
\bibinfo{author}{\bibfnamefont{D.~G.} \bibnamefont{Clarke}},
  \bibinfo{author}{\bibfnamefont{S.~P.} \bibnamefont{Strong}},
  \bibnamefont{and} \bibinfo{author}{\bibfnamefont{P.~W.}
  \bibnamefont{Anderson}}, \bibinfo{journal}{Phys. Rev. Lett.}
  \textbf{\bibinfo{volume}{72}}, \bibinfo{pages}{3218} (\bibinfo{year}{1994}).

\bibitem[{\citenamefont{Schulz}(1996)}]{hjs:prb-rc:96}
\bibinfo{author}{\bibfnamefont{H.~J.} \bibnamefont{Schulz}},
  \bibinfo{journal}{Phys. Rev. B} \textbf{\bibinfo{volume}{53}},
  \bibinfo{pages}{R2959} (\bibinfo{year}{1996}).

\bibitem[{\citenamefont{Renn and Arovas}(1995)}]{qhlinej1}
\bibinfo{author}{\bibfnamefont{S.~R.} \bibnamefont{Renn}} \bibnamefont{and}
  \bibinfo{author}{\bibfnamefont{D.~P.} \bibnamefont{Arovas}},
  \bibinfo{journal}{Phys. Rev. B} \textbf{\bibinfo{volume}{51}},
  \bibinfo{pages}{16832} (\bibinfo{year}{1995}).

\bibitem[{\citenamefont{Kane and Fisher}(1997)}]{qhlinej2}
\bibinfo{author}{\bibfnamefont{C.~L.} \bibnamefont{Kane}} \bibnamefont{and}
  \bibinfo{author}{\bibfnamefont{M.~P.~A.} \bibnamefont{Fisher}},
  \bibinfo{journal}{Phys. Rev. B} \textbf{\bibinfo{volume}{56}},
  \bibinfo{pages}{15231} (\bibinfo{year}{1997}).

\bibitem[{\citenamefont{Mitra and Girvin}(2001)}]{aditi:prb-rc:01}
\bibinfo{author}{\bibfnamefont{A.}~\bibnamefont{Mitra}} \bibnamefont{and}
  \bibinfo{author}{\bibfnamefont{S.~M.} \bibnamefont{Girvin}},
  \bibinfo{journal}{Phys. Rev. B} \textbf{\bibinfo{volume}{64}},
  \bibinfo{pages}{041309(R)} (\bibinfo{year}{2001}).

\bibitem[{\citenamefont{Kollar and Sachdev}(2002)}]{kollar:prb-rc:02}
\bibinfo{author}{\bibfnamefont{M.}~\bibnamefont{Kollar}} \bibnamefont{and}
  \bibinfo{author}{\bibfnamefont{S.}~\bibnamefont{Sachdev}},
  \bibinfo{journal}{Phys. Rev. B} \textbf{\bibinfo{volume}{65}},
  \bibinfo{pages}{121304} (\bibinfo{year}{2002}).

\bibitem[{\citenamefont{Kim and Fradkin}()}]{fradline}
\bibinfo{author}{\bibfnamefont{E.-A.} \bibnamefont{Kim}} \bibnamefont{and}
  \bibinfo{author}{\bibfnamefont{E.}~\bibnamefont{Fradkin}},
  \bibinfo{note}{cond-mat/0205629}.

\bibitem[{\citenamefont{Z\"ulicke}(1999)}]{uz:prl:99}
\bibinfo{author}{\bibfnamefont{U.}~\bibnamefont{Z\"ulicke}},
  \bibinfo{journal}{Phys. Rev. Lett.} \textbf{\bibinfo{volume}{83}},
  \bibinfo{pages}{5330} (\bibinfo{year}{1999}).

\bibitem[{\citenamefont{Naud et~al.}(2000)\citenamefont{Naud, Pryadko, and
  Sondhi}}]{naud:nucl:00}
\bibinfo{author}{\bibfnamefont{J.~D.} \bibnamefont{Naud}},
  \bibinfo{author}{\bibfnamefont{L.~P.} \bibnamefont{Pryadko}},
  \bibnamefont{and} \bibinfo{author}{\bibfnamefont{S.~L.}
  \bibnamefont{Sondhi}}, \bibinfo{journal}{Nucl. Phys. B}
  \textbf{\bibinfo{volume}{565}}, \bibinfo{pages}{572} (\bibinfo{year}{2000}).

\bibitem[{\citenamefont{{von Delft} and Schoeller}(1998)}]{vondelft}
\bibinfo{author}{\bibfnamefont{J.}~\bibnamefont{{von Delft}}} \bibnamefont{and}
  \bibinfo{author}{\bibfnamefont{H.}~\bibnamefont{Schoeller}},
  \bibinfo{journal}{Ann. Phys. (Leipzig)} \textbf{\bibinfo{volume}{7}},
  \bibinfo{pages}{225} (\bibinfo{year}{1998}).

\bibitem[{\citenamefont{Z\"ulicke and MacDonald}(1996)}]{uz:prb:96}
\bibinfo{author}{\bibfnamefont{U.}~\bibnamefont{Z\"ulicke}} \bibnamefont{and}
  \bibinfo{author}{\bibfnamefont{A.~H.} \bibnamefont{MacDonald}},
  \bibinfo{journal}{Phys. Rev. B} \textbf{\bibinfo{volume}{54}},
  \bibinfo{pages}{16813} (\bibinfo{year}{1996}).

\bibitem[{\citenamefont{Alekseev et~al.}(1998)\citenamefont{Alekseev,
Cheianov, and Fr\"ohlich}}]{frohlich}
\bibinfo{author}{\bibfnamefont{A.~Yu.} \bibnamefont{Alekseev}},
  \bibinfo{author}{\bibfnamefont{V.~V.}~\bibnamefont{Cheianov}},
  \bibnamefont{and}
  \bibinfo{author}{\bibfnamefont{J.}~\bibnamefont{Fr\"ohlich}},
  \bibinfo{journal}{Phys. Rev. Lett.} \textbf{\bibinfo{volume}{81}},
  \bibinfo{pages}{3503} (\bibinfo{year}{1998}).

\bibitem[{\citenamefont{Tsvelik}(1995)}]{tsvelik}
\bibinfo{author}{\bibfnamefont{A.~M.} \bibnamefont{Tsvelik}},
  \emph{\bibinfo{title}{Quantum Field Theory in Condensed Matter Physics}}
  (\bibinfo{publisher}{Cambridge U.\ Press},
  \bibinfo{address}{Cambridge, UK}, \bibinfo{year}{1995}).

\bibitem[{\citenamefont{Boese et~al.}(2001)\citenamefont{Boese, Governale,
  Rosch, and Z\"ulicke}}]{uz:prb:01}
\bibinfo{author}{\bibfnamefont{D.}~\bibnamefont{Boese}},
  \bibinfo{author}{\bibfnamefont{M.}~\bibnamefont{Governale}},
  \bibinfo{author}{\bibfnamefont{A.}~\bibnamefont{Rosch}}, \bibnamefont{and}
  \bibinfo{author}{\bibfnamefont{U.}~\bibnamefont{Z\"ulicke}},
  \bibinfo{journal}{Phys. Rev. B} \textbf{\bibinfo{volume}{64}},
  \bibinfo{pages}{085315} (\bibinfo{year}{2001}).

\bibitem[{\citenamefont{Z\"ulicke and Shimshoni}()}]{longv}
\bibinfo{author}{\bibfnamefont{U.}~\bibnamefont{Z\"ulicke}}
\bibnamefont{and}
\bibinfo{author}{\bibfnamefont{E.}~\bibnamefont{Shimshoni}},
  \bibinfo{note}{in preparation}.

\bibitem[{\citenamefont{Ho}(1994)}]{tinlunho}
\bibinfo{author}{\bibfnamefont{T.-L.} \bibnamefont{Ho}},
  \bibinfo{journal}{Phys. Rev. B} \textbf{\bibinfo{volume}{50}},
  \bibinfo{pages}{4524} (\bibinfo{year}{1994}).

\end{thebibliography}

\end{document}